%% file: paper.tex
\documentclass[sigconf,authorversion=true]{acmart}

\AtBeginDocument{%
  }

\acmYear{2023}
\copyrightyear{2023}
\setcopyright{rightsretained}
\acmConference[MobiSys '23]{The 21st Annual International Conference on Mobile Systems, Applications and Services}{June 18--22, 2023}{Helsinki, Finland}
\acmBooktitle{The 21st Annual International Conference on Mobile Systems, Applications and Services (MobiSys '23), June 18--22, 2023, Helsinki, Finland}
\acmPrice{15.00}
\acmDOI{10.1145/3581791.3597510}
\acmISBN{979-8-4007-0110-8/23/06}

\input{packages.tex}

\begin{document}

\title{A Path to Holistic Privacy in Stream Processing Systems}

\author{Mikhail Fomichev}
\affiliation{%
  \institution{Technical University of Darmstadt}
  \country{}
}
\email{mfomichev@seemoo.tu-darmstadt.de}

\renewcommand{\shortauthors}{Mikhail Fomichev}

\begin{abstract}
  \input{section/abstract.tex}
\end{abstract}





\maketitle

\input{section/introduction.tex}
\input{section/related_work.tex}
\input{section/methodology.tex}
\input{section/acknowledgement.tex}
\input{section/conclusion.tex}

\vspace{-2ex}
\bibliographystyle{ACM-Reference-Format}
\balance
\bibliography{bibliography}

\end{document}

%% file: packages.tex
\usepackage{booktabs} 
\usepackage[acronym, nowarn]{glossaries}
\makeglossaries

\usepackage{makecell}
\usepackage{multirow}

\usepackage{diagbox}
\usepackage{caption}
\usepackage[font=footnotesize,labelfont=bf]{caption}

\usepackage{xcolor,colortbl}

\usepackage{xspace}

\usepackage{tikz}

\usepackage{halloweenmath} 

\usepackage{cleveref}

\usepackage{transparent}

\usepackage{adjustbox}

\usepackage{listings, xcolor}

\definecolor{yelcir}{RGB}{255,192,0}
\definecolor{yelrim}{RGB}{188,140,0}
\definecolor{grecir}{RGB}{112,173,71}
\definecolor{grerim}{RGB}{80,126,50}
\definecolor{blucir}{RGB}{91,155,213}
\definecolor{blurim}{RGB}{65,113,156}
\definecolor{salcir}{RGB}{245,176,147}
\definecolor{salrim}{RGB}{237,125,49}

\definecolor{gr}{RGB}{198,196,215} 
\definecolor{pp}{RGB}{255,121,108} 

\input{abbreviations.tex}

%% file: abbreviations.tex
\newacronym{cep}{CEP}{Complex Event Processing}
\newacronym{put}{PUT}{privacy-utility tradeoff}
\newacronym{imu}{IMU}{inertial measurement unit}
\newacronym{iot}{IoT}{Internet of Things}
\newacronym{har}{HAR}{human activity recognition}
\newacronym{adl}{ADL}{activities of daily living}
\newacronym{ppm}{PPM}{privacy-preserving mechanism}
\newacronym{ml}{ML}{machine learning}
\newacronym{spe}{SPE}{stream processing engine}
\newacronym{qos}{QoS}{quality of service}
\newacronym{polp}{PoLP}{principle of least privilege}
\newacronym{dp}{DP}{differential privacy}
\newacronym{sps}{SPS}{stream processing system}
\newacronym{sp}{SP}{Stream Processing}
\newacronym{vae}{VAE}{variational autoencoder}
\newacronym{tee}{TEE}{trusted execution environment}
\newacronym{nlp}{NLP}{natural language processing}
\newacronym{ac}{AC}{access control}

\newacronym{lstm}{LSTM}{long short-term memory}
\newacronym{hmm}{HMM}{hidden Markov model}
\newacronym{dbn}{DBN}{dynamic Bayesian network}

\newacronym{ai}{AI}{artificial intelligence}
\newacronym{hbc}{HBC}{honest-but-curious}
\newacronym{hci}{HCI}{human-computer interaction}

%% file: section/abstract.tex
The massive streams of \gls{iot} data require a timely analysis to retain data usefulness.
\Glspl{sps} enable this task, deriving knowledge from the \gls{iot} data in real-time. 
Such real-time analytics benefits many applications but can also be used to violate user privacy, as the \gls{iot} data collected from users or their vicinity is inherently sensitive. 
In this paper, we present our systematic look into privacy issues arising from the intersection of \glspl{sps} and \gls{iot}, identifying key research challenges towards achieving holistic privacy protection in \glspl{sps} and proposing the solutions. 

%% file: section/introduction.tex

\glsresetall
\vspace{-2ex}
\section{Introduction}
\label{sec:intro}
Rich data \textit{sensed} by \gls{iot} devices enables various applications, ranging from healthcare to autonomous driving~\cite{ding2022roadmap, fomichev2021fastzip}. 
\Glspl{sps} allow \textit{inferring knowledge} from the massive volumes of data produced by \gls{iot} devices on the fly. 
This empowers \textit{real-time analytics} in different domains, like e-commerce and health monitoring, reducing the amount of data that needs to be stored and data staleness. 
Current \glspl{sps} \textit{distribute} their processing resources between sensor, fog, and cloud layers, allowing \glspl{sps} to \textit{lower latency} by making inferences from the data closer towards its source, e.g., utilizing \gls{ai}~\cite{ding2022roadmap}.
Thus, we refer to such \glspl{sps} as \textit{multilayered} and focus on them in this paper. 

Since multilayered \glspl{sps} are recently adopted, their privacy implications have \textit{not} been thoroughly studied~\cite{burkhalter2022privacy}.
Yet, the \gls{sps}'s ability to infer knowledge in real-time from the data, which may be collected from users---and hence be sensitive, makes \glspl{sps} privacy-risky (e.g., repurposable for tracking), thereby they face \textit{compliance} with data protection regulations, like GDPR, and increasing \textit{users' demand} for better privacy~\cite{fomichev2023no}.  
The multilayeredness of \glspl{sps} poses new privacy challenges, e.g., each layer could confront \textit{different privacy threats} and be \textit{governed by various entities}.
These new challenges add up to other hurdles for privacy protection in \glspl{sps}, like their real-timeness requiring \glspl{ppm}, that are applied in an \gls{sps}, to impose minimum overhead. 

In the following, we present our research towards holistic privacy protection in \glspl{sps}, \textit{to draw a balance} between privacy and utility.

%% file: section/related_work.tex

\vspace{-2ex}
\section{Problem Statement: Privacy in SPS}
\label{sec:rwork}
While \glspl{sps} are already used to analyze \gls{iot} data (e.g., health monitoring), the privacy issues caused by such an analysis have \textit{not} been well-understood~\cite{fomichev2023no}.  
This makes it difficult to select, customize, and apply existing \glspl{ppm} in multilayered \glspl{sps}.
Hence, our \textit{first challenge} is to identify critical threats to user privacy, which generically stem from analyzing \gls{iot} data by a multilayered \gls{sps}.  
We look into such privacy threats under an \textit{\gls{hbc}} adversary model, reflecting \glspl{sps} in use, and do \textit{not} consider malicious attackers, who can be addressed by cryptographic- and authentication-based \glspl{ppm} developed for stream processing settings~\cite{burkhalter2022privacy}. 

Balancing privacy and utility, i.e., a \textit{\gls{put}}, is challenging in all domains where \glspl{ppm} are applied~\cite{cheng2022task}. 
In \glspl{sps}, achieving a \gls{put}---our \textit{second challenge}---is even more difficult due to the real-timeness of \glspl{sps}.
Thus, the \gls{put} becomes twofold: (1) it means \textit{balancing quantifiable privacy and utility metrics}, such as reidentification attack's success and accuracy of recognized human activity, while (2) ensuring that \textit{latency} imposed by \glspl{ppm} does not violate real-time constraints.
The low-latency requirement demands the \glspl{ppm} to add minimum overhead, coming on top of other critical \gls{sps}'s needs, i.e., scaling to a large number of data flows as well as generalizing to heterogeneous (\gls{iot}) data~\cite{johnson2020chorus}. 

The aforementioned two challenges make holistic privacy protection in multilayered \glspl{sps} a difficult problem. 
Next, we provide our approach to address this problem in the context of \gls{iot}, highlighting main findings and avenues for future work. 

%% file: section/methodology.tex

\vspace{-2ex}
\section{Contributions and Outlook} 
\label{sec:res-meth}
\noindent
\textbf{Privacy State in \glspl{sps}: Threats and \glspl{ppm}.} 
To address challenges outlined in~\autoref{sec:rwork}, we utilize---for clarity and without the loss of generality in the \gls{iot} context---a \gls{har} application enabled by  \gls{imu} sensors, e.g., an accelerometer and gyroscope, to identify the critical privacy threats in multilayered \glspl{sps} and \glspl{ppm} against them.
Thus, the \gls{sps} workflow is as such: user wearable devices, like a smartwatch, called \textit{data producers}, collect  \gls{imu} data (\textit{sensor layer}) which is transformed into simple events, e.g., ``moving'', (\textit{fog layer}) that are aggregated to infer complex events, like ``exercising '' by the user (\textit{cloud layer}). 
To monitor user's activity, the \gls{har} application, called a \textit{data consumer}, queries nodes on the sensor, fog, or cloud layers. 

\begin{figure}
\centering
  \includegraphics[width=0.77\linewidth]{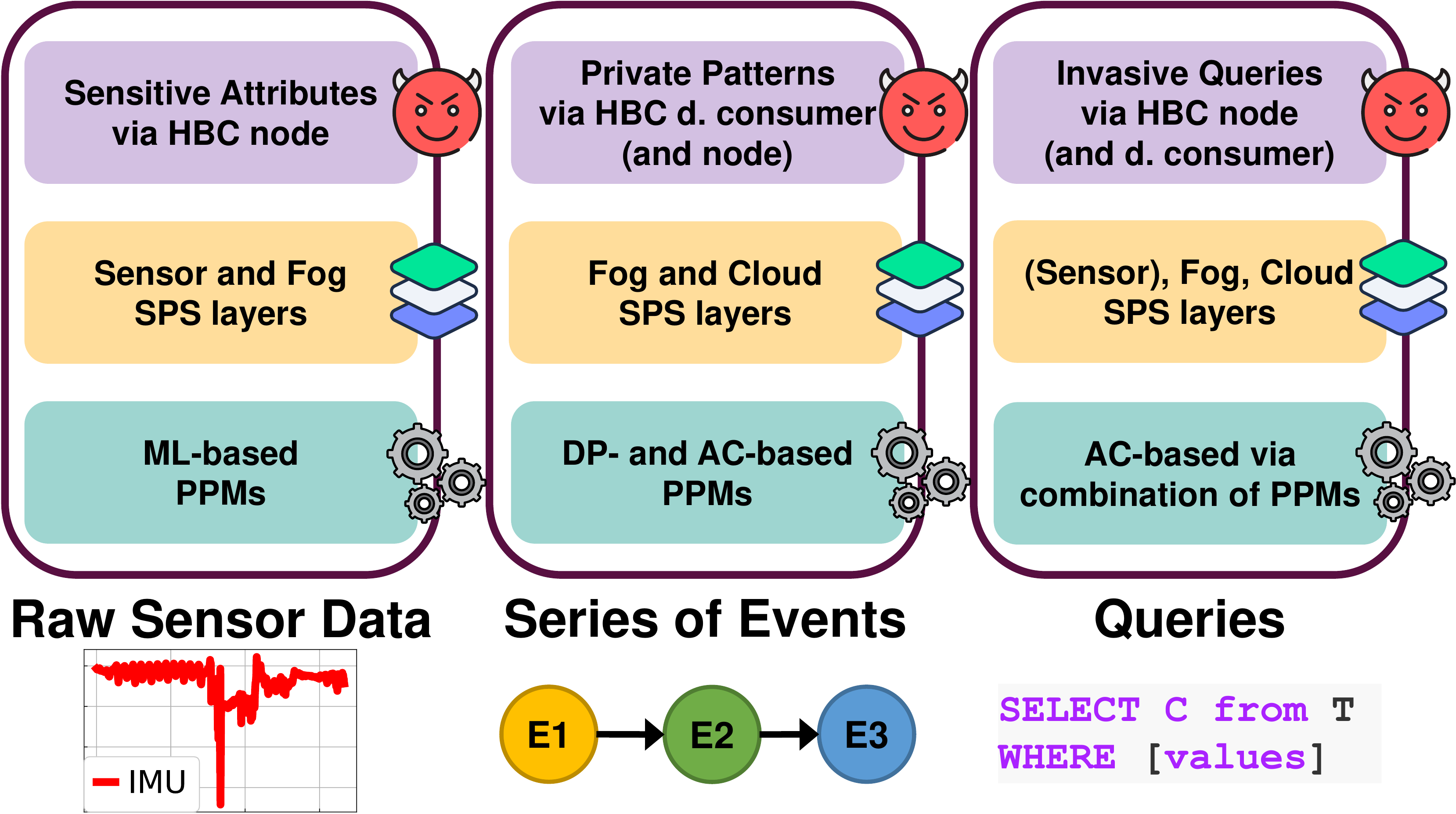}
  \vspace{-2ex}
  \caption{Privacy state overview (i.e., threats and \glspl{ppm}) in multilayered \glspl{sps}.}
  \vspace{-3.5ex}
  \label{fig:approach-overview}
\end{figure}

Based on the \gls{har} use case, we identify (a nonexhaustive list of) critical privacy threats and their conduct in multilayered \glspl{sps} as well as \glspl{ppm} that holistically address such threats~\cite{fomichev2023no} (cf.~\autoref{fig:approach-overview}). 
We find \textit{three pillars} through which user privacy can be violated in \glspl{sps} by the \gls{hbc} adversary: (1) raw sensor data, (2) series of events, and (3) queries. 
Then, for each pillar, we determine a critical privacy threat and its attack vector, i.e., via an \gls{sps} node (e.g., smart hub) or data consumer (e.g., \gls{har} app); specify on each \gls{sps} layer(-s)---sensor, fog, or, cloud---this threat can occur; and propose \glspl{ppm} to address it. 
The revealed threats are: (1) \textit{sensitive attributes}, like inferring user's gender from raw \gls{imu} data, (2) \textit{private patterns}, e.g., detecting user's sensitive events, like ``taking medicine'' from a series of nonsensitive events, and (3) \textit{invasive queries}, which are stealthily crafted to track a user's lifestyle. 
We investigate the applicability of \textit{\gls{ml}}-, \textit{\gls{dp}}-, and \textit{\gls{ac}}-based \glspl{ppm} to tackle these privacy threats. 

\smallskip
\noindent
\textbf{Privacy Policies: A \gls{put} Enabler.} To achieve a \gls{put} in \glspl{sps} (cf. \autoref{sec:rwork}), we suggest using the \glspl{ppm} in a \textit{task-aware fashion}, i.e., protecting data and assets for which users care, like hiding private patterns \textit{only} inside timeframes of their probable occurrence, e.g., ``taking medicine'' by a user around mealtimes. 
This makes the \gls{ppm} application \textit{selective}, focusing on protecting \textit{user-defined} privacy needs while reducing the \gls{ppm} impact on utility. 
We showcase how to utilize a user privacy policy to customize the \gls{dp}-based \gls{ppm} for task-awareness, improving the \gls{put} in \glspl{sps}~\cite{fomichev2023no}.
While these results are promising (and in line with prior findings~\cite{cheng2022task}), we see two \textit{open questions} to achieve mature task-aware privacy protection in \glspl{sps}: (1) is the \textit{formation} of comprehensive user privacy policies and (2) their \textit{translation} into parametrization of different \glspl{ppm} for various \glspl{sps}' use cases (e.g., \gls{har}) to yield the optimal \gls{put}.

\smallskip
\noindent
\textbf{Scrutinizing \glspl{ppm}: Suitability for \glspl{sps}.}
In addition to achieving the composite \gls{put}, as described in~\autoref{sec:rwork}, \glspl{ppm} must also satisfy the \gls{sps} needs for \textit{scalability} and \textit{generalizability} to heterogeneous (\gls{iot}) data.  
We find out that recent \gls{ml}-based \glspl{ppm}, transforming raw \gls{imu} data to hide user's sensitive attributes, like gender, have limited scalability (e.g., a few data flows) and generalizability (e.g., dependency on worn-device location)~\cite{fomichev2023no}---thus more research is required to improve such \glspl{ppm}. 
Moreover, we discover the problem of \textit{interdependency} of sensitive attributes, like concealing the gender alone is not sufficient, as it can be reinferred from, e.g., weight and height attributes. 
Such interdependency prompts \textit{rethinking on how to apply} (\gls{ml}-based) \glspl{ppm} to holistically protect sensitive attributes in \glspl{sps}; we anticipate this problem for the private patterns as well. 

\smallskip
\noindent
\textbf{Combining \glspl{ppm}: Their Interplay.}
Our review of privacy threats in \glspl{sps} shows that different \glspl{ppm} can be used to address the same threat, like \gls{dp}- and \gls{ac}-based \glspl{ppm} to protect private patterns (cf. \autoref{fig:approach-overview}). 
Yet, it remains underexplored \textit{how to combine} the best of two-\gls{ppm} worlds to improve privacy protection.  
In addition, we see that some privacy threats (e.g., invasive queries) are too complex to be tackled by a single \gls{ppm}~\cite{fomichev2023no}. 
These results are backed up by prior findings, such as using \gls{ml} techniques to optimize the amount of noise in the task-aware \gls{dp}-based \gls{ppm}~\cite{cheng2022task} and developing a highly scalable \gls{dp}-based \gls{ppm} by combining three sub\glspl{ppm} complementing each other within the so-called \textit{cooperative architecture}~\cite{johnson2020chorus}.

Currently, privacy solutions like~\cite{johnson2020chorus}, which can serve real-world systems, are scarce. 
Thus, we see a research gap in studying \textit{how to combine different \glspl{ppm}} (e.g., \gls{ml}-, \gls{dp}-, and \gls{ac}-based) to accomplish a more efficient and utility-friendly protection against the critical privacy threats in \glspl{sps} (cf.~\autoref{fig:approach-overview}), as well as exploring the generalizability of such combined \glspl{ppm} for various \glspl{sps}' use cases. 
In a situation where a single \gls{ppm} is applied to address a threat per \gls{sps} layer, we identify another research avenue of effectively \textit{interfacing} different-layer's \glspl{ppm} to work as a whole, boosting the \gls{put}.

\smallskip
\noindent
\textbf{Governance in Multilayered \glspl{sps}.}
The distribution of processing resources between the sensor, fog, and cloud layers, which are not necessarily controlled by the same entity, changes the \textit{trust model} of an \gls{sps} from ``I trust a cloud provider'' to ``I have to trust multiple parties''. 
Such a distributed trust model is still less accepted by users, while its adoption can be enabled via open architectures and higher transparency of resource providers~\cite{ding2022roadmap}.
A new privacy challenge for \glspl{sps} is data governance within such a distributed setting, \textit{ensuring compliance} of different resource providers, that do data processing on sensor, fog, or cloud layers, while \textit{supporting various user privacy behaviors}. 
An exemplary research question here is: ``How to handle the aggregated data collected from multiple (e.g., smart home) users, who have different or even opposing privacy behaviors?''~\cite{zhang2022teo}. 

%% file: section/acknowledgement.tex
\vspace{-2ex}
\section{Acknowledgments}
\label{sec:ack}
This work has been co-funded by the Research Council of Norway as part of the project Parrot (311197) and the Deutsche Forschungsgemeinschaft (DFG) within the Collective Resilient Unattended Smart Things (CRUST) project as part of the DFG Priority Program SPP 2378 -- Resilient Worlds.

%% file: section/conclusion.tex
\glsresetall

\vspace{-2ex}
\section*{Biography}
Mikhail Fomichev obtained his PhD degree in Computer Science, focusing on pairing and authentication for \gls{iot} devices, from TU Darmstadt in 2021.  
Now, as a postdoc at TU Darmstadt, he researches privacy issues stemming from the intersection of \glspl{sps} and the \gls{iot}.

%% file: paper.bbl

\begin{thebibliography}{7}


\ifx \showCODEN    \undefined \def \showCODEN     #1{\unskip}     \fi
\ifx \showDOI      \undefined \def \showDOI       #1{#1}\fi
\ifx \showISBNx    \undefined \def \showISBNx     #1{\unskip}     \fi
\ifx \showISBNxiii \undefined \def \showISBNxiii  #1{\unskip}     \fi
\ifx \showISSN     \undefined \def \showISSN      #1{\unskip}     \fi
\ifx \showLCCN     \undefined \def \showLCCN      #1{\unskip}     \fi
\ifx \shownote     \undefined \def \shownote      #1{#1}          \fi
\ifx \showarticletitle \undefined \def \showarticletitle #1{#1}   \fi
\ifx \showURL      \undefined \def \showURL       {\relax}        \fi
\providecommand\bibfield[2]{#2}
\providecommand\bibinfo[2]{#2}
\providecommand\natexlab[1]{#1}
\providecommand\showeprint[2][]{arXiv:#2}

\bibitem[Burkhalter(2022)]%
        {burkhalter2022privacy}
\bibfield{author}{\bibinfo{person}{Lukas Burkhalter}.}
  \bibinfo{year}{2022}\natexlab{}.
\newblock \emph{\bibinfo{title}{{Privacy-Centric Systems for Stream Data
  Processing}}}.
\newblock \bibinfo{thesistype}{Ph.\,D. Dissertation}. \bibinfo{school}{ETH
  Zurich}.
\newblock


\bibitem[Cheng et~al\mbox{.}(2022)]%
        {cheng2022task}
\bibfield{author}{\bibinfo{person}{Jiangnan Cheng}, \bibinfo{person}{Ao Tang},
  {and} \bibinfo{person}{Sandeep Chinchali}.} \bibinfo{year}{2022}\natexlab{}.
\newblock \showarticletitle{{Task-aware Privacy Preservation for
  Multi-dimensional Data}}. In \bibinfo{booktitle}{\emph{International
  Conference on Machine Learning}}. \bibinfo{pages}{3835--3851}.
\newblock


\bibitem[Ding et~al\mbox{.}(2022)]%
        {ding2022roadmap}
\bibfield{author}{\bibinfo{person}{Aaron~Yi Ding} {et~al\mbox{.}}}
  \bibinfo{year}{2022}\natexlab{}.
\newblock \showarticletitle{{Roadmap for Edge AI: A Dagstuhl Perspective}}.
\newblock \bibinfo{journal}{\emph{ACM SIGCOMM Computer Communication Review}}
  \bibinfo{volume}{52}, \bibinfo{number}{1} (\bibinfo{year}{2022}),
  \bibinfo{pages}{28--33}.
\newblock


\bibitem[Fomichev et~al\mbox{.}(2021)]%
        {fomichev2021fastzip}
\bibfield{author}{\bibinfo{person}{Mikhail Fomichev}, \bibinfo{person}{Julia
  Hesse}, \bibinfo{person}{Lars Almon}, \bibinfo{person}{Timm Lippert},
  \bibinfo{person}{Jun Han}, {and} \bibinfo{person}{Matthias Hollick}.}
  \bibinfo{year}{2021}\natexlab{}.
\newblock \showarticletitle{{FastZIP: Faster and More Secure Zero-interaction
  Pairing}}. In \bibinfo{booktitle}{\emph{Proceedings of the 19th Annual
  International Conference on Mobile Systems, Applications, and Services}}.
  \bibinfo{pages}{440--452}.
\newblock


\bibitem[Fomichev et~al\mbox{.}(2023)]%
        {fomichev2023no}
\bibfield{author}{\bibinfo{person}{Mikhail Fomichev}, \bibinfo{person}{Manisha
  Luthra}, \bibinfo{person}{Maik Benndorf}, {and} \bibinfo{person}{Pratyush
  Agnihotri}.} \bibinfo{year}{2023}\natexlab{}.
\newblock \showarticletitle{{No One Size (PPM) Fits All: Towards Privacy in
  Stream Processing Systems}}.
\newblock \bibinfo{journal}{\emph{arXiv preprint arXiv:2305.00871}}
  (\bibinfo{year}{2023}).
\newblock


\bibitem[Johnson et~al\mbox{.}(2020)]%
        {johnson2020chorus}
\bibfield{author}{\bibinfo{person}{Noah Johnson}, \bibinfo{person}{Joseph~P
  Near}, \bibinfo{person}{Joseph~M Hellerstein}, {and} \bibinfo{person}{Dawn
  Song}.} \bibinfo{year}{2020}\natexlab{}.
\newblock \showarticletitle{{Chorus: A Programming Framework for Building
  Scalable Differential Privacy Mechanisms}}. In \bibinfo{booktitle}{\emph{2020
  IEEE European Symposium on Security and Privacy (EuroS\&P)}}.
  \bibinfo{pages}{535--551}.
\newblock


\bibitem[Zhang et~al\mbox{.}(2022)]%
        {zhang2022teo}
\bibfield{author}{\bibinfo{person}{Han Zhang}, \bibinfo{person}{Yuvraj
  Agarwal}, {and} \bibinfo{person}{Matt Fredrikson}.}
  \bibinfo{year}{2022}\natexlab{}.
\newblock \showarticletitle{{TEO: Ephemeral Ownership for IoT Devices to
  Provide Granular Data Control}}. In \bibinfo{booktitle}{\emph{Proceedings of
  the 20th Annual International Conference on Mobile Systems, Applications and
  Services}}. \bibinfo{pages}{302--315}.
\newblock


\end{thebibliography}
